\documentclass[10pt,conference]{IEEEtran}
\IEEEoverridecommandlockouts
\usepackage{cite}
\usepackage[square,numbers]{natbib}
\usepackage{amsmath,amssymb,amsfonts}
\usepackage{algorithmic}
\usepackage{graphicx}
\usepackage{textcomp}
\usepackage{xcolor}
\setcitestyle{numbers}
\usepackage{hyperref}
\def\BibTeX{{\rm B\kern-.05em{\sc i\kern-.025em b}\kern-.08em
    T\kern-.1667em\lower.7ex\hbox{E}\kern-.125emX}}
\begin{document}

\title{Human-In-The-Loop Software Development Agents: Challenges and Future Directions}

\makeatletter
\newcommand{\newlineauthors}{%
  \end{@IEEEauthorhalign}\hfill\mbox{}\par
  \mbox{}\hfill\begin{@IEEEauthorhalign}
}
\makeatother

\author{\IEEEauthorblockN{Jirat Pasuksmit\IEEEauthorrefmark{1},
Wannita Takerngsaksiri\IEEEauthorrefmark{2}, Patanamon Thongtanunam\IEEEauthorrefmark{3}, Chakkrit Tantithamthavorn\IEEEauthorrefmark{2},\\ 
Ruixiong Zhang\IEEEauthorrefmark{1},
Shiyan Wang\IEEEauthorrefmark{1},
Fan Jiang\IEEEauthorrefmark{1},
Jing Li\IEEEauthorrefmark{1},
Evan Cook\IEEEauthorrefmark{1},
Kun Chen\IEEEauthorrefmark{1}, 
Ming Wu\IEEEauthorrefmark{1}}
\IEEEauthorblockA{\IEEEauthorrefmark{1}Atlassian, Australia.}
\IEEEauthorblockA{\IEEEauthorrefmark{2}Monash University, Australia.}
\IEEEauthorblockA{\IEEEauthorrefmark{3}The University of Melbourne, Australia.}

}


\maketitle

\begin{abstract}
Multi-agent LLM-driven systems for software development are rapidly gaining traction, offering new opportunities to enhance productivity. 
At Atlassian, we deployed Human-in-the-Loop Software Development Agents to resolve Jira work items and evaluated the generated code quality using functional correctness testing and GPT-based similarity scoring.
This paper highlights two major challenges: the high computational costs of unit testing and the variability in LLM-based evaluations.
We also propose future research directions to improve evaluation frameworks for Human-In-The-Loop software development tools.

\end{abstract}


\section{Introduction}
At Atlassian, we are deploying the Human-In-The-Loop LLM-based Software Development Agents Framework (HULA)~\cite{takerngsaksiri2024human} to enhance engineer productivity by generating code to resolve Jira work items.
In summary, HULA allows software engineers to guide LLM agents in planning and coding tasks while ensuring humans retain full control to review and refine outputs at each step.
To evaluate HULA’s performance, we implemented functional correctness testing with the SWE-bench dataset~\cite{jimenez2023swe, sweverify2024} and similarity scoring using GPT-based Large Language Models (LLM-judge).
These evaluations revealed two key challenges: the high computational costs of functional correctness testing and the variability in LLM-based scoring, which complicates detecting small incremental improvements.
Addressing these challenges requires innovative evaluation frameworks that extend beyond traditional unit testing and leverage the capabilities of LLMs.
This paper outlines these limitations, explores potential solutions, and proposes future research directions to advance evaluation methodologies for AI-assisted software development tools.


\section{Challenges}
\textbf{Scalability Limitations of Functional Correctness Testing:}
HULA’s initial evaluation relied on unit testing with the SWE-bench dataset~\cite{jimenez2023swe, sweverify2024} to verify its ability to solve fundamental programming problems. 
During early development, the SWE-bench unit testing framework served as a critical guardrail metric for assessing functional correctness against selected Github issues and Pull-Requests (PR).
Currently, HULA is achieving a 37.2\% resolution rate on the SWE-bench Verified dataset (n=500)~\cite{sweverify2024}.

However, this approach presented three key challenges. 
First, building an evaluation dataset with unit test cases is labor intensive as Jira work items may not mapped exclusively to a unit test.
Second, the unit testing only reflects the binary result of ``pass" or ``fail", not ``how close the generated code is to being functionally correct."
Third, running these tests is computationally expensive and time-consuming, requiring complex environment setups and code execution.
Future research could explore unit test case generation for functionality correctness testing.
This would allow us to focus our computation power on the unit test cases that are highly relevant to our use cases.
With a higher number of test cases, we will also be able to report functional correctness on a finer-grain scale (e.g., the generated code passed 50\% of the available test cases).

\textbf{Fluctuations in GPT-based Similarity Scoring:}
To ensure HULA can address Jira work items in industrial settings like human engineers do, we deployed an LLM-judge metric to evaluate the alignment of AI-generated code with human-written code.
This approach was chosen for its faster execution time and the advantage of requiring no project setup~\cite{tong-zhang-2024-codejudge}, unlike unit testing.
Therefore, our team uses GPT-4 and GPT-4o as LLM judges to evaluate the similarity between AI-generated code and human-written code at the pull request level using a 1-4 scoring scale.
To interpret the results, we aggregated the scores into ``low similarity" (1-2) and ``high similarity" (3-4) to mitigate fluctuations in LLM output."
We validated the LLM judge and observed strong agreement with human evaluations (correlation coefficient: 0.7, n=203) and its ability to represent unit testing on the SWE-bench Verified dataset (F1-score: 0.67, n=500).
Currently, this LLM-judge metric is being used to draw an experimental conclusion for HULA development.
HULA can generate code that achieved a high similarity score (i.e., scored 3 or 4) for 38.5\% of work items in our internal evaluation set.

Despite its benefits, LLM-based evaluations often produce fluctuating outcomes, limiting the ability to reliably assess small (e.g., 1–2\%) incremental improvements in HULA’s capabilities.
Addressing this issue is critical for making informed technical decisions during the iterative development of HULA.
Future work could focus on creating a more stable and deterministic evaluation framework specific for a Human-In-The-Loop software development agents.
Such a framework could also consider other dimensions beyond functional correctness and code similarity (e.g., \cite{cotroneo2024automating, wang2022recode}).

\setcitestyle{numbers}
\bibliographystyle{ACM-Reference-Format}
\bibliography{main}
\vspace{12pt}

\end{document}